\documentclass[aps,pra,showpacs,floatfix,twocolumn]{revtex4}
\usepackage{graphicx}
\usepackage{amsmath}
\usepackage{amssymb}
\usepackage{hyperref}
\usepackage{latexsym}
\usepackage{epsfig}


\begin{document}

\title{Ultracold polar molecules near quantum degeneracy}

\author{S. Ospelkaus$^1$, K.-K. Ni$^1$, M. H. G. de Miranda$^1$, B. Neyenhuis$^1$, D. Wang$^1$, S. Kotochigova$^2$, P. S. Julienne$^3$, D. S. Jin$^1$, and J. Ye$^1$}
\affiliation{$^1$JILA, National Institute of Standards and Technology and University of Colorado\\
Department of Physics, University of Colorado, Boulder, CO 80309-0440, USA
\\ $^2$Physics Department, Temple University, Philadelphia, PA 19122-6082, USA
\\ $^3$Joint Quantum Institute, NIST and University of Maryland, Gaithersburg, MD 20899-8423, USA}

\begin{abstract}
We report the creation and characterization of a near quantum-degenerate gas of polar $^{40}$K-$^{87}$Rb molecules in their absolute rovibrational ground state. Starting from weakly bound heteronuclear KRb Feshbach molecules, we implement precise control of the molecular electronic, vibrational, and rotational degrees of freedom with phase-coherent laser fields. In particular, we coherently transfer these weakly bound molecules across a 125 THz frequency gap in a single step into the absolute rovibrational ground state of the electronic ground potential.  Phase coherence between lasers involved in the transfer process is ensured by referencing the lasers to two single components of a phase-stabilized optical frequency comb. Using these methods, we prepare a dense gas of $4\cdot10^4$ polar molecules at a temperature below $400$\,nK. This fermionic molecular ensemble is close to quantum degeneracy and can be characterized by a degeneracy parameter of $T/T_F=3$. We have measured the molecular polarizability in an optical dipole trap where the trap lifetime gives clues to interesting ultracold chemical processes. Given the large measured dipole moment of the KRb molecules of $0.5$\, Debye, the study of quantum degenerate molecular gases interacting via strong dipolar interactions is now within experimental reach.
\end{abstract}


\maketitle

\section{Introduction}
Ultracold molecular quantum systems promise to open exciting new scientific directions ranging from ultra-cold chemistry~\cite{cold_chemistry1, cold_chemistry2} and precision measurements~\cite{precision1, precision2, precision3} to the realization of novel quantum many-body systems~\cite{many_body}. In particular, polar molecules, which possess permanent electric dipole moments, have very unique perspectives as systems with long-range and anisotropic interactions~\cite{anisotropic}. This is in sharp contrast to the short-range contact interaction dominating atomic Bose-Einstein condensates and degenerate Fermi gases. The strong dipole-dipole interaction between polar molecules adds intriguing and entirely new aspects to the physics of ultracold quantum matter. By means of external static electric fields, dipoles can be oriented along the field axis. This allows for precise control of interactions between the constituents of the quantum gas, resulting in novel dynamics such as anisotropic collision resonances~\cite{collisions1, collisions2}. When these dipoles are lined up in a lattice geometry provided by optical potentials, the long-range interaction introduces complex quantum dynamics that can be precisely controlled via the field strength and orientation. These types of control of interactions have been the basis for numerous exciting theoretical proposals ranging from quantum phase transitions~\cite{many_body} to novel systems for quantum information processing~\cite{quantumcomp,qc1,qc2} and quantum control with external magnetic and electric fields~\cite{quantum_control}.

However, the observation of dipolar interactions in a molecular gas is challenging and requires the dipolar interaction energy of the molecular gas to be at least comparable to its kinetic energy. This requirement can be fulfilled only with an ultracold, high-density gas of polar molecules with a reasonably large electric dipole moment, namely a polar molecular gas near quantum degeneracy.

Experimental approaches towards the realization of a quantum degenerate gas of ultracold polar molecules have generally followed two different strategies. The first approach is to develop experimental techniques for direct cooling of ground-state polar molecules. In recent years, research in this direction has resulted in powerful techniques such as buffer gas cooling~\cite{buffergas}, Stark deceleration~\cite{Starkdeceleration1,Starkdeceleration2}, or velocity filtering~\cite{velocity}. However, direct cooling strategies have typically been restricted to the millikelvin temperature range and very low phase space densities that are a trillion times away from quantum degeneracy. For further advances along this direction, laser cooling of molecules will have to be introduced~\cite{lasercoolingm}. The second approach is to make use of the powerful cooling techniques for atoms and subsequently convert dual-species atom pairs to deeply bound polar molecules. The challenges in this context are to make this conversion process efficient and to effectively remove the binding energy of the deeply bound molecules without heating the resulting molecular ensemble. In brief, the desirable high phase space density of the initial atomic ensemble has to be preserved during the conversion process. Photoassociation of ultracold atoms provides an efficient way of removing the binding energy of molecules by means of  spontaneously emitted photons~\cite{photoassociation, limit1, limit2}. However, the spontaneous processes result in a large spread of population among many continuum and bound molecular states, leading to significant dilution of the initial atomic phase space density. As a result,  typical molecular phase space densities at the end of this process are comparable to the current best results from direct cooling of ground state molecules.

Recently, we reported a fully coherent approach to efficiently convert a near quantum-degenerate atomic gas into a near quantum-degenerate polar molecular gas in the absolute rovibrational ground state~\cite{Ni2008}. A heteronuclear quantum gas mixture of fermionic $^{40}$K and bosonic $^{87}$Rb is first efficiently converted into weakly bound Feshbach molecules in the vicinity of a Fano-Feshbach resonance~\cite{Feshbach, exp2,exp1}. This initial conversion process is efficient and coherent. The system is now prepared in a single molecular quantum state, even though it is a highly excited vibrational level. Following this initial association process, we then use highly phase-coherent laser fields to precisely and efficiently transfer populations between internal states of specific electronic, vibrational and rotational degrees of freedom~\cite{Ospelkaus2008, coherent_transfer}. For our specific goal, we convert the weakly bound Feshbach molecules into the absolute ground state, namely the zero vibration and zero rotation quantum level in the singlet electronic ground-state molecular potential. Given the large energy gap between the near-threshold Feshbach molecules and the rovibrational ground state, phase coherence between the lasers used in the transfer process can only be ensured by referencing to individual components of a phase-stabilized optical frequency comb~\cite{CundiffYe}. In this article we report the further refinement of our technique after the initial demonstration~\cite{Ni2008} and the preliminary characterization of some of the properties for absolute ground-state polar molecules near quantum degeneracy.

\section{Experimental Techniques and Challenges}
The starting point for our approach is a near quantum degenerate gas mixture of fermionic $^{40}$K atoms and bosonic $^{87}$Rb atoms confined in a single beam far-off-resonance optical dipole trap. A DC magnetic-field is ramped across a Fano-Feshbach resonance at 546.7 G to associate pairs of K and Rb atoms into extremely weakly bound heteronuclear KRb molecules~\cite{exp1,exp2}. This association process converts up to 25\% of the atoms into weakly bound molecules. We create up to $5\cdot 10^4$ heteronuclear Feshbach molecules with a binding energy of $h\cdot 230$\,kHz, where $h$ is Planck's constant. The resulting trapped gas of fermionic near-threhold molecules has a density of $n=5\cdot10^{11}/$cm$^3$ and an expansion energy of $T=400(15)$\,nK. The gas of Feshbach molecules is thus close to quantum degeneracy with  $T/T_F\approx 3$. Here $T_F$ is the Fermi temperature of the molecules.

\begin{figure}
    \includegraphics[width=7cm,angle=-90]{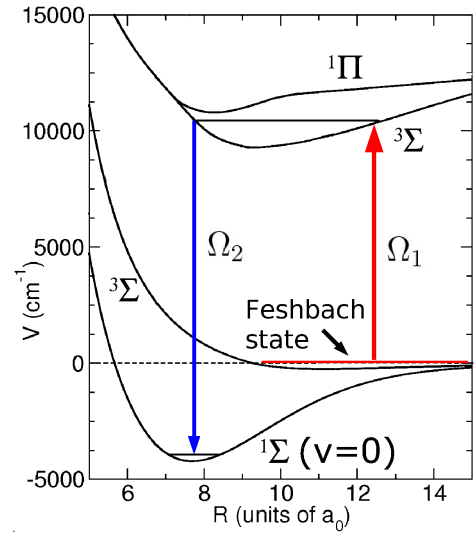}
  \caption{Sketch  of the KRb electronic ground  and excited molecular potentials involved in the two-photon coherent state transfer from the Feshbach molecules to the rovibrational ground state $X^1{\Sigma_0}(v=0)$. We choose the vibrational level v=23 of the $2^3\Sigma_1$ excited state molecular potential as the intermediate coupling state. This state has favorable Franck-Condon factors to both the initial Feshbach state and the $X^1\Sigma_0$ rovibrational ground state. In addition, its proximity to the bottom of the $1^1{\Pi_1}$ excited state molecular potential ensures the necessary singlet triplet mixing to convert predominantly triplet character Feshbach molecules to  singlet rovibrational ground state molecules. Laser 1 connecting the Feshbach state to the intermediate excited state is operating at $970$\,nm. Laser 2, connecting the intermediate state to the rovibrational ground state, operates at $690$\,nm. The Raman laser system therefore bridges the binding energy difference between Feshbach and rovibrational ground state molecules of $\Delta E_B\approx h\cdot125 \, $THz }
\label{fig:scheme}
\end{figure}

The next challenge is to transfer the weakly bound molecules into their absolute ground state. We emphasize that our goal is transfer to a single internal quantum state without any heat dissipations to external motional states. Hence, a fully coherent conversion process is key. The conversion efficiency is also of paramount importance in order to preserve the initial high phase-space density. As shown in Fig.~\ref{fig:scheme}, a coherent Raman transfer process involving an electronically excited intermediate state can be envisioned for the transfer process between the initial and final vibration levels. Due to the vastly mismatched vibrational wave-functions of the weakly bound and the absolute ground-state molecules, it was initially regarded as nearly impossible to find a suitable intermediate state that can provide reasonable resonance strengths for both the upward and downward transitions. In fact, a theory proposal was made to employ a train of two-color, phase-coherent pulses that would allow coherent accumulations of a pump-dump process to implement a fully coherent molecular conversion process \cite{Avi2007}. The properly time-delayed pump-dump sequence takes advantage of the excited vibration dynamics to enhance the transition probability while the coherent accumulation gradually selects a single final state. Systematic and detailed single photon spectroscopy ensued, connecting the initial Feshbach state to specific electronically excited states with a CW laser referenced to an optical frequency comb. An intense theory-experiment collaboration soon led us to the realization that we can vastly reduce the complexity of the problem by using a single stationary intermediate state instead of dynamic wave-packets.  As indicated in Fig.~\ref{fig:scheme}, the key is to find an intermediate state where the inner turning point closely matches the Condon point for the downward transition (labeled by the field $\Omega_2$) and the outer turning point lies close to the Condon point for the up-going transition (labeled by $\Omega_1$). In doing so, the chosen intermediate state provides favorable Franck-Condon factors to both the initial weakly bound Feshbach state and the rovibrational ground state. In addition, the intermediate state lies in proximity to the crossing of  $2^3\Sigma$ and  $1^1\Pi$ excited potentials, which ensures the necessary singlet-triplet mixing to couple Feshbach molecules with a predominantly triplet character to the singlet rovibrational ground state. We believe that such a suitable intermediate state can always be found for any bi-alkali heteronuclear molecular system, thus making our approach a universally applicable technique for the production of different kinds of bi-alkali polar molecules.

A Raman laser system consists of two different-colored CW lasers of reliable mutual phase coherence. The difference frequency between the two Raman lasers needs to bridge an energy gap of 125 THz, corresponding to the binding energy of the absolute rovibrational ground state of the KRb molecule. This challenging requirement can only be met by referencing and individually phase-locking the two lasers to a femtosecond optical frequency comb~\cite{CundiffYe}, which itself is phase locked to an ultra-stable Nd:YAG laser at 1064\,nm. We use the dark resonance technique~\cite{dark_res} to search and precisely determine energy positions of the rovibrational ground state. In particular, we determine the binding energy of the rovibrational ground state $X^1\Sigma_0\,(v = J = 0)$  to be 125.319703(1)\,THz~\cite{Ni2008}.  The relevant single-photon transition strength is also determined in the dark resonance two-photon spectroscopy.

\begin{figure}
    \includegraphics[width=\columnwidth]{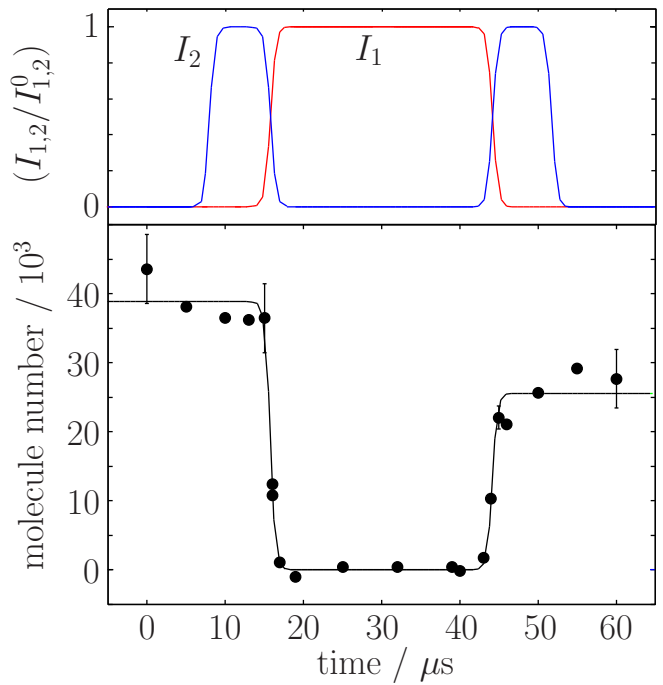}
  \caption{Time evolution of the coherent two-photon transfer (STIRAP) from Feshbach molecules to the absolute rovibrational ground state $X^1\Sigma_0\,(v=0)$. (a) Counterintuitive STIRAP pulse sequence, here $I_1$ and $I_2$ are the intensities of laser 1 and laser 2, respectively (see Figure~\ref{fig:scheme}). (b) Measured population in the initial Feshbach state during the STIRAP pulse sequence. Starting with about $4\cdot 10^4$  Feshbach molecules, the molecules  are coherently transferred to the rovibrational ground state $X^1\Sigma_0\,(v=0)$   by the first pulse sequence (t =15 to 20\,$\mu$s).  The rovibrational ground state molecules are  invisible to the detection light. Reversing the pulse sequence, $X^1\Sigma_0\,(v=0)$ molecules are converted back to weakly bound  Feshbach molecules (t =45\,$\mu$s to t=$50$\,$\mu$s).}
\label{fig:time_series}
\end{figure}

For population transfer, we use a single step of STIRAP (Stimulated Raman Adiabatic Passage)~\cite{STIRAP} to convert Feshbach molecules into rovibrational ground state $X^1\Sigma_0\,(v = J = 0)$ molecules, as shown schematically in Fig.~\ref{fig:scheme}.
Figure~\ref{fig:time_series} displays the coherent two-photon transfer (STIRAP) process from Feshbach molecules to the absolute rovibrational ground state. The top panel depicts the counterintuitive STIRAP pulse sequence, where the field of $\Omega_2$ is turned on first to establish coherence between the intermediate and the final target states. As $\Omega_2$ ramps down, field $\Omega_1$ ramps up, transferring the population adiabatically from the initial Feshbach state to the final ground-state target. To monitor the population transfer process, we measure the population in the initial Feshbach state during the STIRAP pulse sequence by direct resonant absorption imaging~\cite{exp2,hetFesh1}. The rovibrational ground state molecules are invisible to the detection light. The population evolution is shown in the bottom panel of Fig.~\ref{fig:time_series}. About $4\cdot 10^4$ Feshbach molecules are coherently transferred to the rovibrational ground state $X^1\Sigma_0\,(v=0)$ by the first pulse sequence (t = 15 to 20\,$\mu$s), resulting in near-zero signal for detection of Feshbach molecules after the first pulse. To demonstrate that these molecules are indeed hidden in the absolute ground state, we reverse the pulse sequence and convert the $X^1\Sigma_0\,(v=0, J=0)$ molecules back to weakly bound  Feshbach molecules (t =40\,$\mu$s to t=$45$\,$\mu$s), allowing for direct detection again.

\section{Characterization of the transfer process}

\begin{figure}
    \includegraphics[width=\columnwidth]{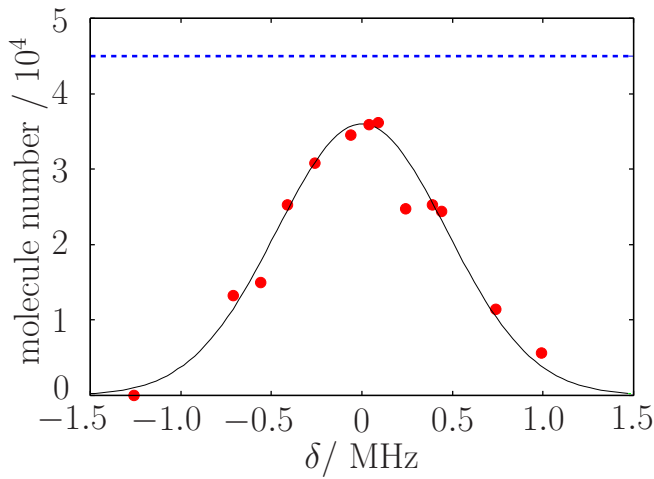}
  \caption{STIRAP lineshape. Plotted is the  number of Feshbach molecules returning after a full round-trip STIRAP sequence as a function of the two-photon Raman laser detuning $\delta$. Starting with $4.5\cdot 10^4$ Feshbach molecules (blue dashed line), we recover about $3.6\cdot 10^4$ Feshbach molecules after a round-trip STIRAP. This corresponds to a transfer efficiency of 80\% for the round-trip transfer suggesting a creation efficiency of ground state polar molecules of  90\%.}
\label{fig:lineshape}
\end{figure}

\begin{figure}
    \includegraphics[width=\columnwidth]{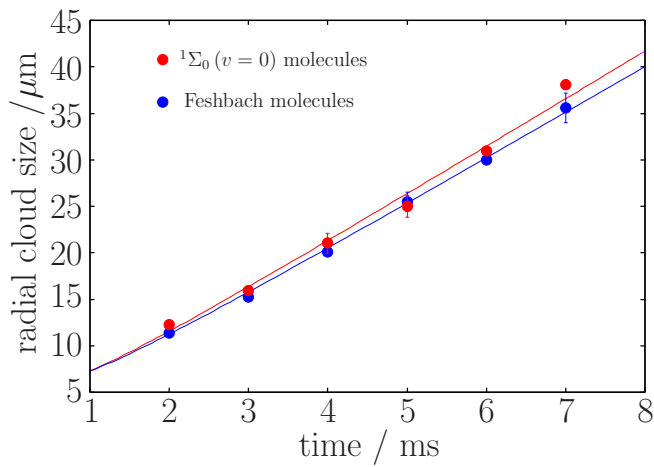}
  \caption{Comparison of the kinetic energy of the Feshbach molecules before STIRAP transfer (blue circles) and after a round-trip STIRAP process (red circles). The latter can be interpreted as an upper limit on the kinetic energy of the  rovibrational ground state molecules. The temperature of both clouds is extracted by time-of-flight expansion analysis. We extract $T=400(15)$\,nK for the Feshbach molecules and $T=430(20)$\,nK for rovibrational ground state molecules. The analysis suggests that the transfer process does not cause any noticeable heating on the molecules.}
\label{fig:temperature}
\end{figure}

The fraction of the returned molecules gives  the efficiency of the round-trip STIRAP process. To optimize the STIRAP process, we precisely scan the frequency difference between the two lasers so as to search for the two-photon Raman resonance.   Figure~\ref{fig:lineshape} shows a typical STIRAP lineshape, where the number of Feshbach molecules returning after a full round-trip STIRAP sequence is measured as a function of the two-photon Raman laser detuning $\delta$. Starting with $4.5\cdot 10^4$ Feshbach molecules (indicated by the blue dashed line in the figure), we recover about $3.6\cdot 10^4$ Feshbach molecules after a round-trip STIRAP. This corresponds to a transfer efficiency of 80\% for the round-trip transfer, suggesting a transfer efficiency to ground-state polar molecules of 90\%. To prove that this transfer process is fully coherent and does not result in any motional heating of the molecular sample, we compare the kinetic energy of the absolute ground-state molecules versus the kinetic energy of the initial Feshbach molecules. This comparison is accomplished by time-of-flight expansion measurement on the Feshbach molecules before the STIRAP transfer and after a round-trip STIRAP process. Figure~\ref{fig:temperature} documents the time-of-flight expansion measurement results, with blue circles corresponding to the Feshbach molecules before STIRAP transfer and the red circles for those after a round-trip STIRAP process. Clearly, the latter molecules are expanding at nearly the same kinetic energy as the former ones, indicating that the rovibrational ground-state molecules are translationally as cold as the initial population. In fact, by extracting the temperature from the expansion analysis, we find $T=400(15)$\,nK for the Feshbach molecules and $T=430(20)$\,nK for rovibrational ground state molecules. We have thus experimentally established that our transfer process is highly efficient and fully coherent, allowing 90\% of the initial phase-space density to be transferred into the absolute ground state.

\section{Properties of KRb ground state polar molecules}
The rovibrational level structure of the entire $X^1\Sigma_0$ electronic ground potential can be mapped out using the frequency comb-assisted two-photon dark resonance spectroscopy. In Ref.~\cite{Ni2008}, we measured the rotational constant for the $v=0$ manifold. The dipole moment of the KRb rovibrational ground state can be measured via Stark spectroscopy performed on the two-photon dark resonance. Since the initial Feshbach state has no appreciable electric dipole moment, the frequency shift of the two-photon Raman resonance under an external DC electric field arises solely from the level shift of the absolute ground state. Since we already know the rotational structure, the dipole moment can thus be precisely determined to be $d=0.566(17)$\, Debye (see Ref.~\cite{Ni2008}). In the near future, we also plan to map out the hyperfine structure~\cite{hyperfine} of the KRb absolute ground-state using high-resolution dark resonance spectroscopy.

\begin{figure}
    \includegraphics[width=\columnwidth]{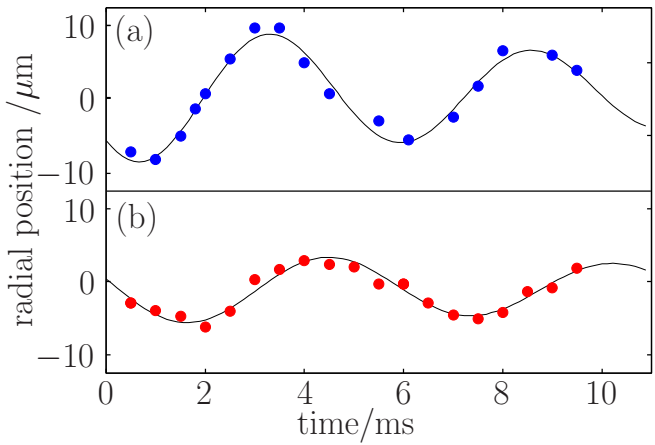}
  \caption{Comparison of the radial slosh of Feshbach molecules (a) vs. ground state molecules (b) in the optical dipole trap at $\lambda=1090\,$nm. The slosh in the trap is excited by perturbing the optical potential for 0.5\,ms and measured after $2\,ms$ of expansion. We extract a trap frequency of $2\pi\cdot 190(3)$\,Hz for Feshbach molecules (a) as compared to a trap frequency of $2\pi\cdot 175(5)\,$Hz for ground state molecules. From the frequency ratio, we derive a preliminary polarizability ratio for of $\alpha_{X^1\Sigma_0} / \alpha_{\mathrm{Feshbach}}=\omega_{X^1\Sigma_0}^2/\omega_{\mathrm{Feshbach}}^2=0.85(5)$ at $\lambda=1090$\,nm.
}
\label{fig:slosh}
\end{figure}

\begin{figure}
    \includegraphics[width=\columnwidth]{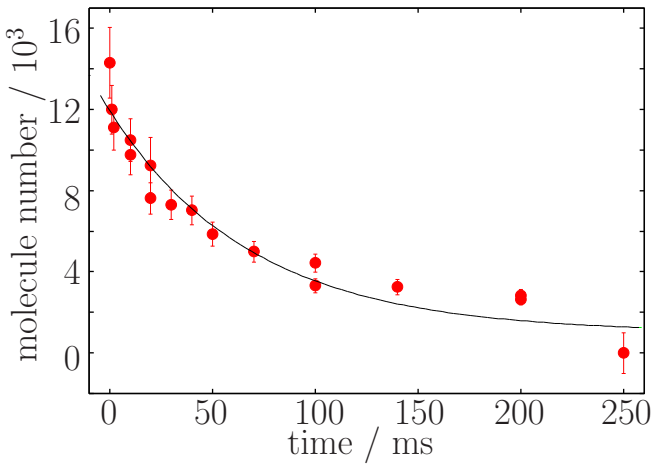}
  \caption{Lifetime of ground-state polar molecules in the optical dipole trap. We extract $\tau=70(8)$\,ms. }
\label{fig:lifetime}
\end{figure}

Another interesting property to measure for these ground-state molecules is their AC polarizability. In fact, the existence of the permanent dipole moment allows us to align these molecules under an external DC electric field of a few kilovolts per centimeter. As these molecules are confined inside an optical dipole trap, we can vary the polarization of the trapping light with respect to the DC electric field and measure the anisotropic property of the polarizability~\cite{Dulieu08}. We have now taken the first step in measuring the average value of the AC polarizability. Theoretically, the AC polarizability of the $X^1\Sigma_0\,v=0$ ground state of the KRb molecule was first investigated in~\cite{alignment}. Experimentally, the AC polarizability can be determined from the oscillation frequencies of trapped molecules provided we know the trapping light intensity.  Figure~\ref{fig:slosh} shows a preliminary comparison of the radial slosh of Feshbach molecules and ground-state molecules in the optical dipole trap at $\lambda=1090\,$nm. The slosh in the trap is excited by perturbing the optical potential for 0.5\,ms.  We observe a pronounced difference between the trap frequencies for Feshbach molecules and ground state molecules. We extract a trap frequency of $2\pi\cdot 190(3)$\,Hz for Feshbach molecules as compared to a trap frequency of $2\pi\cdot 175(5)\,$Hz for ground-state molecules. From the frequency ratio, we derive a polarizability ratio for of $\alpha_{X^1\Sigma_0} / \alpha_{\mathrm{Feshbach}}=\omega_{X^1\Sigma_0}^2/\omega_{\mathrm{Feshbach}}^2=0.85(5)$ at $\lambda=1090$\,nm. The polarizability of the Feshbach molecules is simply the sum of the Rb and K atomic polarizabilities~\cite{polarize} $\alpha_{\mathrm{Feshbach}}=5.7\cdot 10^{-5}$\, MHz/(W/cm$^2$).

One important property of v=0 molecules is their expected long lifetime, which will allow further cooling to quantum degeneracy. The background pressure of our vacuum chamber and the blackbody excitation rate of the rovibrational transitions both support a lifetime longer than a few tens of seconds. The light scattering by the trapping beam should also be fairly weak, and we do not expect it to limit the lifetime below 10 s. The lifetime could be limited by inelastic molecule-molecule and atom-molecule collisions. If the rovibrational ground state $^{40}$K$^{87}$Rb molecules are created in a single hyperfine state, then the Fermi nature of these molecules will preclude their collisions at this ultracold temperature. A possible loss mechanism is the collision of a K atom with a KRb molecule, resulting in a chemical reaction producing a K$_2$ molecule and a Rb atom. The potential minimum of K$_2$ molecules~\cite{K} is located below that of KRb~\cite{KRb}, so that this process is energetically allowed. By removing most of the K atoms, we have measured a typical lifetime of $\tau=70(8)$\,ms for the ground-state polar molecules sitting in the optical dipole trap. Figure~\ref{fig:lifetime} shows one such measurement. When we purposely allow K atoms to remain in the same optical trap and when their number is similar to that of the KRb molecules, we observe a decrease of the trap lifetime to below 10 ms. On the other hand, the potential minimum of the Rb$_2$ molecules is above that of KRb, and introducing Rb atoms to the trapped KRb molecules does not seem to affect the trap lifetime. At this time, we are still searching for the reason behind the 70 ms limit for the lifetime. Planned measurements in the future include the purification of the KRb sample to a single hyperfine state to prevent possible KRb-KRb collisions and the measurement of incoherent scattering of ground-state molecules due to the trapping light.

\section{Conclusions}
The successful production of a high phase-space density ultracold polar molecules has now prepared us to explore an exciting range of scientific topics including simulation of quantum phase transitions, quantum information processing, precision measurement, and novel collisions and chemical reactions at ultralow energies. The field of ultracold molecules is poised to explode with many exciting new results in the near future and we are  eager to be part of this endeavor.

We thank A. Pe'er and J. Zirbel for their technical contributions to the work reported here. Funding support is provided by NSF, NIST, AFOSR, and ARO. S.O. acknowledges support from the Alexander von Humboldt Foundation, K.-K.N. and B.N. from NSF, and M.H.G.de.M. from the CAPES/Fulbright.

\end{document}